\documentclass[aps,twocolumn,nofootinbib,showpacs, prd]{revtex4}
\usepackage{graphicx}
\usepackage{amsfonts}
\usepackage{amssymb}
\usepackage{amsbsy}
\usepackage{amsmath}
\usepackage{latexsym}
\usepackage{natbib}
\usepackage{bm}
\usepackage{subfigure}
\usepackage{color}
\usepackage{float}
\usepackage{comment}

\def\ce{\mathrm{ce}}
\def\se{\mathrm{se}}
\def\sgn{\mathrm{sgn}}

\def\k{\mathbf{k}}

\def\x{\mathbf{x}}
\def\y{\mathbf{y}}

\def\k{\mathrm{k}}

\def\H{\mathcal{H}}

\def\lstar{l_{\star}}

\begin{document}

\title{Emergence of spacetime discreteness in Wightman function from 
polymer
quantization of scalar field
}

\author{Golam Mortuza Hossain}
\email{ghossain@iiserkol.ac.in}

\author{Gopal Sardar}
\email{gopal1109@iiserkol.ac.in}

\affiliation{ Department of Physical Sciences, 
Indian Institute of Science Education and Research Kolkata,
Mohanpur - 741 246, WB, India }
 
\pacs{04.62.+v, 04.60.Pp}

\date{\today}

\begin{abstract}

The Wightman function \emph{i.e.} vacuum two-point function, for a massless 
free scalar field in Fock quantization is inversely proportional to the 
invariant distance squared between the corresponding spacetime points. Naturally 
it diverges when these two points are taken to be infinitesimally close to each 
other. Using a combination of analytical and numerical methods, we show that the 
Wightman function is bounded from above in polymer quantization of scalar field. 
The bounded value of the Wightman function is governed by the polymer scale and 
its bounded nature can be viewed as if the spacetime has a zero-point length. 
This emergence of effective discreteness in spacetime appears from polymer 
quantization of matter field alone as the geometry is treated classically. It 
also demonstrates that the polymer quantization of matter field itself can 
capture certain aspects of the effective discrete geometry. We discuss the 
implications of the result on the response function of a Unruh-DeWitt 
detector that depends on the properties of the Wightman function.

\end{abstract}

\maketitle

\section{Introduction}

The vacuum two-point function which is usually known as the \emph{Wightman 
function} \cite{peskin1995,kaku1993quantum,Birrell1984quantum,Ford:1997hb}, 
plays an important role in disseminating the implications of the quantum field 
theory. As an example, it is often employed to see whether the causality is 
preserved by the field in a given quantum field theory. In addition, the 
Wightman function has been used as a probe to understand the structure of the 
underlying spacetime  
\cite{Blau:2010fh,Kapustin:1999ci,Mercuri:2010xz,Raasakka:2016uyk}. One way to 
visualize this aspect is to note that the Wightman function for a massless free 
scalar field in Fock quantization is inversely proportional to the invariant 
distance squared between the corresponding spacetime points. This relation then 
may be used in reverse to conclude about the nature of the \emph{effective} 
spacetime distance between two given points, as experienced by the field, if one 
knows the corresponding Wightman function. Therefore, if one computes the 
Wightman function in a theory which aims to modify physics near Planck scale 
then one could use it to see whether the properties of such a Wightman function 
indicate any Plank scale alteration to the effective spacetime distance. For 
example, one could ask whether the notion of \emph{zero-point length} can emerge 
from the properties of the Wightman function.

Besides, the Wightman function is also known to play an important role in 
realizing aspects of certain quantum phenomena in curved spacetime such as the 
\emph{Unruh effect} 
\cite{Birrell1984quantum,Crispino:2007eb,Takagi01031986,Unruh:1976db}. In 
particular, the response function of a Unruh-Dewitt detector depends on the 
structure of the \emph{pole} of the Wightman function. In the context of polymer 
quantization, it has been argued that Unruh effect could disappear 
\cite{Hossain:2014fma,Hossain:2015xqa,Hossain:2016klt} due to the presence of a 
new length scale. The polymer quantization 
\cite{Ashtekar:2002sn,Halvorson-2004-35} is a canonical quantization method 
which is used in \emph{Loop Quantum gravity} 
\cite{Ashtekar:2004eh,Rovelli2004quantum,Thiemann2007modern}. Some key aspects 
of polymer quantization are known to differ from Schrodinger quantization when 
one applies it to a mechanical system. Consequently, in polymer quantization 
of the field one expects to find some results which are different from the 
Fock quantization. Therefore, it is imperative to study the properties 
of the Wightman function in the context of polymer quantization and some 
qualitative aspects of it are already noted in 
\cite{Hossain:2015xqa,Hossain:2016klt}. A complete analytical computation of the 
Wightman function in polymer quantization turns out to be very difficult. 
Therefore, in order to understand the behaviour of the Wightman function for a 
wide range of spacetime intervals, in this article we employ a combination of 
both analytical and numerical techniques in the context of polymer quantization 
of scalar matter field in a classical flat geometry.

This paper is organized as follows. In the section 
\ref{Sec:MasslessScalarField}, we consider a massless free scalar field in 
the canonical framework. Subsequently, we review the general derivation of the 
Wightman function in the canonical approach and derive the same for Fock 
quantization in the section \ref{Sec:Fock}. In the section \ref{Sec:Polymer}, we 
briefly discuss some key results from polymer quantization which are relevant 
for computation of the Wightman function. Then, we  analytically compute the 
Wightman function in the asymptotic regions for the timelike and the spacelike 
intervals separately. Subsequently, we employ numerical techniques to compute 
it for a wide range of spacetime intervals. Using both analytical and numerical 
computations, we show that unlike in Fock quantization, the Wightman function is 
bounded from above in polymer quantization. Its bounded nature is governed by 
the new polymer length scale and this property can be viewed as if the spacetime 
has a notion of zero-point length.

\section{Massless scalar field}
\label{Sec:MasslessScalarField}

In order to study the properties of the two-point function, we consider here a 
massless free scalar field $\Phi(x)$ in Minkowski spacetime. The dynamics of 
such a scalar field is described by the action 
\begin{equation}
 \label{ScalarActionMinkowski}
 S_{\Phi} = \int d^4x \left[ - \frac{1}{2} \sqrt{-\eta} \eta^{\mu\nu}  
 \partial_{\mu} \Phi(x)  \partial_{\nu} \Phi(x) \right] ~,
\end{equation}
where the invariant distance element can be expressed as
\begin{equation}\label{MinkowskiMetric}
ds^2 = \eta_{\mu\nu} dx^{\mu} dx^{\nu} = -dt^2 + q_{ab} d\x^a d\x^b ~.
\end{equation}
Here we use the \emph{natural units} $c=\hbar=1$ and the metric is
$\eta_{\mu\nu} = diag(-1,1,1,1)$. In this article, for quantization of the 
matter field, we aim to apply polymer quantization method which is a canonical 
approach. Therefore, we need to compute the Hamiltonian associated with the 
scalar field action (\ref{ScalarActionMinkowski}). By choosing \emph{spatial} 
hyper-surfaces, each of which is labeled by given time $t$, one can express the 
corresponding scalar field Hamiltonian as
\begin{equation}\label{SFHamGen}
H_{\Phi}  =  \int d^3\x \left[ \frac{\Pi^2}{2\sqrt{q}} +
\frac{\sqrt{q}}{2} q^{ab} \partial_a\Phi \partial_b\Phi
\right] ~,
\end{equation}
where $q_{ab}$ is the spatial metric and $q$ is its \emph{determinant}. The 
Poisson bracket between the field $\Phi = \Phi(t,\x)$ and the conjugate field 
momentum $\Pi = \Pi(t,\x)$ is given by
\begin{equation}\label{PositionSpacePB}
\{\Phi(t,\x), \Pi(t,\y)\} = \delta^{3}(\x-\y) ~.
\end{equation}

\subsection{Fourier modes}

In this article we follow the polymer quantization approach as suggested in 
\cite{Hossain:2010eb} where one performs explicit Fourier transformation of 
the field to express it as a set of decoupled simple harmonic oscillators 
\emph{before} quantization. In particular, one defines the Fourier modes for the 
scalar field and its conjugate momentum as
\begin{equation}\label{FourierModesDef}
\Phi = \frac{1}{\sqrt{V}} \sum_{\k} \tilde{\phi}_{\k}(t) e^{i \k\cdot\x} ,~
\Pi  = \frac{1}{\sqrt{V}} \sum_{\k} \sqrt{q} ~\tilde{\pi}_{\k}(t) 
e^{i \k\cdot\x},
\end{equation}
where $V=\int d^3\x \sqrt{q}$ is the spatial volume. In Minkowski spacetime,
due to the non-compactness of the spatial hyper-surfaces, the spatial volume 
will diverge. Therefore, using a fiducial box of finite volume, one can avoid 
to have such a divergent quantity in the intermediate steps. In such 
case, Kronecker delta and Dirac delta are expressed as $\int d^3\x \sqrt{q} 
~e^{i (\k-\k')\cdot \x} = V \delta_{\k,\k'}$ and $\sum_{\k} e^{i \k\cdot 
(\x-\y)} = V \delta^3 (\x-\y)/\sqrt{q}$. 

\subsection{Hamiltonian density of the modes}

Using the expressions of Kronecker delta and Dirac delta, the Hamiltonian 
(\ref{SFHamGen}) can be expressed fully in terms of Fourier modes 
(\ref{FourierModesDef}). In particular, the field 
Hamiltonian (\ref{SFHamGen}) can be expressed as 
$H_{\Phi} = \sum_{\k} \H_{\k}$, where the Hamiltonian density for the $\k^{th}$ 
mode is
\begin{equation}\label{SFHamFourierMinkowski}
\H_{\k} = \frac{1}{2} \tilde{\pi}_{-\k} \tilde{\pi}_{\k} +
\frac{1}{2} |\k|^2 \tilde{\phi}_{-\k}\tilde{\phi}_{\k}  ~.
\end{equation}
The non-vanishing Poisson brackets between the modes are given by
\begin{equation}\label{Minkowski:MomentumSpacePB}
\{\tilde{\phi}_{\k}, \tilde{\pi}_{-\k'}\} = \delta_{\k,\k'} ~.
\end{equation}
The Fourier modes of field $\tilde{\phi}_{\k}$ and its conjugate momenta
$\tilde{\pi}_{\k}$ are in general \emph{complex-valued} functions. Therefore, 
each complex-valued mode has two independent \emph{real-valued} modes. 
For a real-valued scalar field $\Phi$, one can impose a reality condition to 
redefine the complex-valued modes $\tilde{\phi}_{\k}$ and $\tilde{\pi}_{\k}$ in 
terms of the real-valued modes, say, $\phi_{\k}$ and $\pi_{\k}$ respectively. In 
terms of these real-valued mode functions, the corresponding Hamiltonian density 
and the Poisson brackets can be expressed as
\begin{equation}\label{SFHamFourierMinkowskiReal}
\H_{\k} = \frac{1}{2} \pi_{\k}^2 + \frac{1}{2} |\k|^2 \phi_{\k}^2
~~~;~~~ \{\phi_{\k},\pi_{\k'}\} =  \delta_{\k,\k'} ~.
\end{equation}
The equation (\ref{SFHamFourierMinkowskiReal}) represents a standard system of 
decoupled harmonic oscillators. Clearly, a free scalar field can be viewed as 
a composition of infinitely many decoupled harmonic oscillators.

\subsection{General form of Wightman function}

The standard Fock quantization of the scalar field is achieved by quantizing 
these simple harmonic oscillators using Schrodinger quantization as if each of 
these modes were a mechanical system. Analogously, we shall use the polymer 
quantization for these modes in order to perform polymer quantization of the 
scalar field.

In general, we may express the energy spectrum of these quantum oscillators as 
$\hat{\H}_{\k}|n_{\k}\rangle = E_n^{(\k)} |n_{\k}\rangle$ where the energy 
eigenvalues $E_n^{(\k)}$ of the $\k^{th}$ oscillator corresponds to the 
energy eigenstates $|n_{\k}\rangle$. The corresponding vacuum state for the 
field can then be expressed as $|0\rangle=\Pi_{\k}\otimes |0_\k\rangle$. 
Therefore, the general form of the vacuum two-point function which is also 
known as the Wightman function, can be expressed as 
\begin{equation}
\label{MinkowskiTwoPointDef}
G(x,x') \equiv \langle 0|\hat{\Phi}(x) \hat{\Phi}(x')|0\rangle
= \langle 0|\hat{\Phi}(t,\x) \hat{\Phi}(t',\x')|0\rangle ~.
\end{equation}
In terms of the Fourier modes (\ref{FourierModesDef}), one can express the 
Wightman function (\ref{MinkowskiTwoPointDef}) as
\begin{equation}
\label{MinkowskiTwoPointDef2}
G(x,x') = \frac{1}{V} \sum_{\k} D_{\k}(t,t') ~e^{i {\k}\cdot(\x-\x')} ,
\end{equation}
where the matrix element can be written as
\begin{equation}\label{DkDefinition}
D_{\k}(t,t') = 
\langle 0_{\k}| e^{i\hat{\H}_{\k}t} \hat{\phi}_{\k} e^{-i\hat{\H}_{\k}t}
e^{i\hat{\H}_{\k}t'} \hat{\phi}_{\k} e^{-i\hat{\H}_{\k}t'}
|0_{\k}\rangle.
\end{equation}
We may expand the state $\hat{\phi}_{\k} |0_{\k}\rangle$ in the complete basis 
of the energy eigenstates as $\hat{\phi}_{\k}|0_{\k}\rangle = \sum_{n} c_n 
|n_{\k}\rangle$. Subsequently, by using the energy spectrum, we can express the 
matrix element as
\begin{equation}
\label{DkFunctionGeneral}
D_{\k}(t-t') \equiv D_{\k}(t,t') = \sum_{n} |c_n|^2 e^{-i\Delta E_n (t-t')},
\end{equation}
where the energy gaps $\Delta E_n \equiv E_n^{(\k)} - E_0^{(\k)}$ and the
coefficients $c_n = \langle n_{\k}| \hat{\phi}_{\k} |0_{\k}\rangle$.

Thanks to the chosen definition of Fourier modes (\ref{FourierModesDef}), the  
Hamiltonian density and Poisson brackets of the modes are already 
independent of the fiducial volume. The remaining fiducial volume dependence 
in the expression of the Wightman function (\ref{MinkowskiTwoPointDef2}) can be 
removed by taking the limit $V \to \infty$ such that the summation becomes an 
integral. In other words, we can remove the fiducial volume, by essentially 
replacing the sum $\frac{1}{V} \sum_{\k}$ by an integration $\int 
\frac{d^3\k}{(2\pi)^3}$. The Wightman function (\ref{MinkowskiTwoPointDef2})  
then becomes
\begin{equation}
\label{GeneralTwoPointIntegralDef}
G(x,x') = \int \frac{d^3\k}{(2\pi)^3} ~  D_{\k}(t,t') 
~e^{i {\k}\cdot(\x-\x')} ~.
\end{equation}
The matrix element of $D_{\k}(t-t')$ depends on $\k$ through its magnitude
for both Fock quantization and the polymer quantization. Therefore, in order
to evaluate the integration (\ref{GeneralTwoPointIntegralDef}), we use 
\emph{polar coordinates} in momentum space as following
\begin{equation}
\label{KGPropagator}
G(x,x') =  \int \frac{k^2 dk}{4\pi^2} D_{k}(\Delta t) 
\int \sin\theta d\theta ~e^{i k |\Delta \x| \cos\theta }  ,
\end{equation}
where $k = |\k|$, $\Delta \x = \x-\x'$ and $\Delta t = t-t'$. 
By performing the integration over the angle $\theta$, one can express the 
Wightman function as
\begin{equation}
\label{KGPropagatorDiffPM}
G(x,x') = G_{+} - G_{-} ~,
\end{equation}
where 
\begin{equation}
\label{GPMDefinition}
G_{\pm} =  \frac{i}{4\pi^2|\Delta\x|} \int dk ~k D_{k}(\Delta t) 
~e^{\mp i k |\Delta \x|} ~.
\end{equation}

Using the equations (\ref{KGPropagatorDiffPM}) and (\ref{GPMDefinition}), we 
note that the Wightman function depends only on the \emph{magnitude} of the 
spatial separation $\Delta \x$. On the other hand it is sensitive also to the 
\emph{sign} of the temporal separation $\Delta t$. An important consequence of 
these dependences is that the vacuum expectation value of the commutator 
bracket $\langle 0|[\hat{\Phi}(x), \hat{\Phi}(x')]|0\rangle = G(x,x')-G(x',x) = 
0$ when temporal separation $\Delta t = 0$ for the given two spacetime points 
$x$, $x'$. For a spacelike separation in Minkowski spacetime, one can always 
choose a frame where the temporal separation $\Delta t$ vanishes. These aspects 
together imply that no causal communication is possible between any two 
spacelike events.

In general, the integrals in the equation (\ref{GPMDefinition}) are not 
convergent. Therefore, one needs to employ some form of regularization 
techniques to render them finite. Here we use the standard prescription where 
one introduces a non-oscillatory regulator term in the integral as follows
\begin{equation}
\label{GPMDefRegulated}
G_{\pm}^{\delta} =  \frac{i}{4\pi^2|\Delta\x|} \int dk ~k D_{k}(\Delta t) 
~e^{\mp i k |\Delta \x|} ~e^{- \delta ~k} ~.
\end{equation}
Here $\delta$ is taken to be a positive parameter such that 
$\lim_{\delta\to 0} G_{\pm}^{\delta} = G_{\pm}$.

\section{Fock quantization}\label{Sec:Fock}

As we have mentioned earlier that the Fock quantization is achieved by 
quantizing the modes (\ref{SFHamFourierMinkowskiReal}) using Schrodinger 
quantization. In particular, there one seeks to represent the elementary 
commutator bracket $\left[\hat{\phi}_{\k},\hat{\pi}_{\k}\right] = i$ for the 
given mode on the Hilbert space. It is then straightforward to compute 
the energy spectrum for the $\k^{th}$ mode which is given by $E_n^{(\k)} = 
(n+\frac{1}{2})|\k|$. The corresponding coefficients are 
$c_n=\delta_{1,n}/\sqrt{2|\k|}$ and the energy gaps are $\Delta E_n=n|\k|$. The 
Wightman function (\ref{KGPropagatorDiffPM}) for the Fock quantization then 
becomes
\begin{equation}\label{WightmanFockRegulated}
G^{\delta}(x,x')=\frac{1}{4\pi^2\left[ -(\Delta t -i\delta)^2 
+ |\Delta \x|^2\right]} ~,
\end{equation}
where the parameter $\delta$ is introduced as the regulator of the integral. If 
we remove the regulator by taking the limit $\delta\to 0$ then the Wightman 
function (\ref{WightmanFockRegulated}) becomes
\begin{equation}\label{wightmanFock}
G(x,x')=\frac{1}{4\pi^2 \Delta x^2} ~,
\end{equation}
where $\Delta x^2 =  -\Delta t^2+|\Delta \x|^2$ is the usual invariant distance 
interval in the Minkowski spacetime.

\section{Polymer quantization}\label{Sec:Polymer}

We now briefly review some relevant results from polymer quantization for 
the computation of the Wightman function. The polymer quantization is a 
canonical quantization method which is used in loop quantum gravity. Apart from 
the Planck constant $\hbar$, it comes with a new dimension-full parameter, say 
$l_{\star}$. In full quantum gravity, this length scale would correspond to the 
Planck length. However, in this article we treat the background geometry as 
classical and we apply polymer quantization only for the matter field.

In polymer quantization it is shown that the exact energy spectrum of the 
$\k^{th}$ harmonic oscillator can be expressed in terms of the  \emph{Mathieu 
characteristic value functions} $A_n$, $B_n$ as \cite{Hossain:2010eb}  
\begin{equation}
 \label{EigenValueMCFRelation}
 \frac{E_{\k}^{2n}}{|\k|} = \frac{1}{4g} + \frac{g}{2} ~A_n(g)  ~,~
 \frac{E_{\k}^{2n+1}}{|\k|} = \frac{1}{4g} + \frac{g}{2} B_{n+1}(g) ~,
\end{equation}
where $n\ge0$ and $g = |\k|~\lstar$ is a 
\emph{dimension-less} parameter. The corresponding energy eigenstates are 
$\psi_{2n}(v) = \ce_n(1/4g^2,v)/\sqrt{\pi}$ and 
$\psi_{2n+1}(v) = \se_{n+1}(1/4g^2,v)/\sqrt{\pi}$ where 
$v = \pi_{\k} \sqrt{\lstar} + \pi/2 $.
These functions $\ce_n$ and $\se_n$ are solutions to \emph{Mathieu
equations} which are known as elliptic cosine and sine functions
respectively \cite{Abramowitz1964handbook}. We should mention here
that the superselection rules are employed to arrive at these 
$\pi$-periodic and $\pi$-antiperiodic states in $v$. Without imposition of any 
such superselection rules, some statistical features of the system are known to 
be ill-defined \cite{Barbero:2013lia}.

In polymer quantization, the coefficients  $c_{4n+3} = i \sqrt{\lstar} 
\int_0^{2\pi} \psi_{4n+3} \partial_v\psi_{0} dv$ are non-vanishing for all 
positive integer $n$  whereas in Fock quantization only \emph{one} $c_n$ is 
non-vanishing. Using the asymptotic expressions of $A_n$ and $B_n$, it is shown 
that the energy spectrum (\ref{EigenValueMCFRelation}) reduces to regular 
harmonic oscillator energy spectrum along with perturbative corrections in the 
small $g$ limit \cite{Hossain:2010eb}. In particular, the relevant energy gaps 
for \emph{sub-Planckian modes} (\emph{i.e.} $g\ll 1$) can be expressed as
\begin{equation}\label{SmallgDeltaE4n+3}
\frac{\Delta E_{4n+3}}{|\k|} = (2n+1) - \frac{(4n+3)^2-1}{16} g  + 
\mathcal{O} \left( g^2 \right) ~,
\end{equation}
for $n \ge 0$. On the other hand, for \emph{super-Planckian} modes (\emph{i.e.} 
$g\gg1$), one can approximate the energy gaps as
\begin{equation} \label{LargegDeltaE4n+3}
 \frac{\Delta E_{4n+3} }{|\k|}  =  2(n+1)^2 g + 
  \mathcal{O}\left(\frac{1}{g^3}\right) ,
\end{equation}
for $n \ge 0$. One may note that the form of the energy gaps 
(\ref{LargegDeltaE4n+3}) for super-Planckian modes significantly differ from the 
corresponding energy gaps obtained from Fock quantization.

Using asymptotic behavior of the Mathieu functions in small $g$ limit, one can 
approximate the coefficients as
\begin{equation}
\label{SmallgC4n+3}
  c_3 = \frac{i}{\sqrt{2|\k|}} \left[1 
  +\mathcal{O}\left(g\right) \right] ~,~
  \frac{c_{4n+3}}{c_{3}} = \mathcal{O}\left(g^n\right),
\end{equation}
where $n >0$. On the other hand, for the limit $g\gg1$, the coefficients can be 
expressed as
\begin{equation}
  \label{LargegC4n+3}
  c_{3} =  i\sqrt{\frac{g}{2|\k|}} \left[\frac{1}{4g^2} +
  \mathcal{O}\left(\frac{1}{g^6}\right) \right],
  \frac{c_{4n+3}}{c_{3}} = \mathcal{O}\left(\frac{1}{g^{2n}}\right),
\end{equation}
for $n > 0$. Using the equations (\ref{SmallgC4n+3}) and (\ref{LargegC4n+3}), 
it is clear that in polymer quantization one can approximate the matrix elements
as
\begin{equation}\label{DkApproxPolymer}
D_\k(\Delta t) \simeq |c_3|^2e^{-i\Delta E_3 \Delta t}
\equiv |c_k|^2e^{-i\Delta E_k \Delta t}
 ~,
\end{equation}
for both asymptotic regions.

\subsection{Sub-Planckian and super-Planckian contributions to Wightman 
function}

Unlike in the case of Fock quantization, the exact analytical computation of 
the Wightman function for all possible spacetime intervals does not seem to be 
possible in polymer quantization. However, it is possible to perform analytic 
computation for asymptotic regions and some qualitative aspects of it has 
already been noted in \cite{Hossain:2015xqa,Hossain:2016klt}. Asymptotic 
expressions of the Wightman function also becomes very useful for checking the 
consistency of its numerically evaluated values.

From the equations (\ref{GPMDefinition}) and (\ref{DkApproxPolymer}), we note 
that the expression of the Wightman function is determined by the expressions 
of $|c_k|^2 \equiv |c_3|^2$ and energy gap $\Delta E_k \equiv \Delta E_3$.
Further, we note that the asymptotic expressions of $|c_k|^2$ and $\Delta E_k $ 
are different for the \emph{sub-Planckian} and the \emph{super-Planckian} 
modes. By choosing a pivotal wave-number $k_0$ such that 
$g_0 = k_0 \lstar = \mathcal{O}(1)$, we may approximate their expressions 
for sub-Planckian modes ($g<g_0$) as
\begin{equation}\label{DECKSubPlanckian}
\Delta E_{k} \simeq k\left(1 - \tfrac{1}{2} ~\lstar k \right ) ~,~
|c_k|^2 \simeq \frac{1}{2k} \left(1 - 2\delta_{c} ~\lstar k\right) ~,
\end{equation}
where $\delta_{c}$ can be estimated numerically and its value is
$\delta_{c} \approx 0.25$. On the other hand, for super-Planckian modes 
($g>g_0$) we may approximate them as
\begin{equation}\label{DECKSuperPlanckian}
\Delta E_{k} \simeq 2~\lstar ~k^2   ~,~ 
|c_k|^2 \simeq \frac{1}{32~\lstar^3 ~k^4} ~.
\end{equation}
We note here that by requiring both asymptotic expressions for energy gaps 
(\ref{DECKSubPlanckian}) and (\ref{DECKSuperPlanckian}) to match at $g=g_0$
we could obtain an equation of the form $1 - \tfrac{1}{2} g_0 = 2 g_0$
which leads to the value $g_0=0.40$. On the other hand, one could also 
match the coefficient $|c_k|^2$ using numerical value of $\delta_c$ 
and obtain the value $g_0 \approx 0.43$. We should mention here that we have 
obtained asymptotic expressions by neglecting higher order terms. Therefore, 
the given value of $g_0$ here should be considered as a crude estimate. 
However, it is clear that in order to evaluate the Wightman function one should 
consider the contributions from sub-Planckian and super-Planckian modes 
separately as
\begin{equation}
 G= G_{sub} + G_{super} ~,
\end{equation}
where $G_{sub}$ and $G_{super}$ contain the contributions from the 
sub-Planckian and the super-Planckian modes respectively.

\subsection{Asymptotic forms of Wightman function for timelike intervals}

We may note that the energy gap $\Delta E_{k}$ is modified in polymer 
quantization compared to the Fock quantization. Together with the equation 
(\ref{GPMDefinition}), it then implies that the temporal separation contributes 
differently to the Wightman function than a spacelike separation. Therefore, in 
order to compute asymptotic expressions of the Wightman function, we treat 
the timelike and spacelike intervals separately. Further, we also do not 
consider null-like intervals here as it diverges identically for Fock 
quantization and would not be amenable under numerical computations. Using the 
identity $e^{-x}= e^x \sum_{m=0}^{\infty}(-2x)^m/m!$, the Wightman function 
$G(x,x')=G_{+}-G_{-}$(\ref{KGPropagatorDiffPM}) can also be expressed 
as a series of the form
\begin{equation}\label{KGPSeries}
 G=\sum_{m=1}^{\infty}\frac{(-2i|\Delta \x|)^{m-1}}{2\pi^2 m!}\int_0^{\infty}
 dk~ k^{m+1}D_{k}(\Delta t)e^{i k|\Delta \x|}~.
\end{equation}
For a timelike separation, we can always choose the reference frame where 
$\Delta \x =0$. The equation (\ref{KGPSeries}) then reduces to the form
\begin{equation}\label{WightmanTimelikeSimplified0}
G  = \frac{1}{2\pi^2}\int dk~k^2 |c_k|^2 e^{-i\Delta E_k \Delta t} ~.
\end{equation}
By defining a variable $u=\Delta E_{k}\Delta t$, the regulated expression of 
the Wightman function can be written as
\begin{equation}\label{WightmanTimelikeSimplified}
G^{\epsilon} = \int du~h(u,\Delta t) e^{-iu} e^{-\epsilon u} ~,
\end{equation}
such that $\lim_{\epsilon\to0} G^{\epsilon} = G$. The function $h(u,\Delta t)$ 
is given by
\begin{equation}\label{hfunctionGeneralTimelike}
h(u,\Delta t) = \frac{k^2 |c_k|^2 }{2\pi^2} \frac{dk}{du} ~.
\end{equation}
The sub-Planckian contribution to the Wightman function can be expressed as
\begin{equation}\label{GsubTimelike}
 G^{\epsilon}_{sub} = 
\int_{0}^{u_0} du~h_{sub}(u,\Delta t)~e^{-u(\epsilon+i)} ~,
\end{equation}
where  $u_0=\Delta E_{k0}\Delta t = \tilde{g}_0(\Delta t/\lstar)$ and 
$\tilde{g}_0 = g_0(1-g_0/2)$. The function $h_{sub}$ can be expressed as
\begin{equation}\label{hfunctionTimelikeSubPlanckian}
h_{sub}(u,\Delta t) = \frac{1}{4\pi^2} \frac{u}{\Delta t^2}
\left[1 + \frac{\delta_{1}~ \lstar u}{\Delta t} \right]  ~,
\end{equation}
where $\delta_{1} = \tfrac{3}{2} -2\delta_c$. On the other hand, the 
super-Planckian contribution to the Wightman function can be expressed as
\begin{equation}\label{GsuperTimelike}
 G^{\epsilon}_{super} = 
\int_{u_0}^{\infty} du~h_{super}(u,\Delta t)~e^{-u(\epsilon+i)} ~,
\end{equation}
where the function $h_{super}$ is given by
\begin{equation}\label{hfunctionTimelikeSubPlanckian}
h_{super}(u,\Delta t) = \frac{1}{64\pi^2~\lstar^2} 
\sqrt{ \frac{\Delta t}{2\lstar} } ~u^{-3/2} ~.
\end{equation}
It turns out that the analytic evaluation of the equations (\ref{GsubTimelike}) 
and (\ref{GsuperTimelike}) are possible for asymptotically large or small values 
of the interval $\Delta t$.

\subsubsection{Long-distance Wightman function}

In the long distance domain where $\Delta t \gg \lstar$, the limit of 
integration $u_0 \sim (\Delta t/\lstar) \gg 1$. Further, we choose the 
regulator $\epsilon$ such that $(\epsilon u_0) \gg 1$ is also satisfied. This 
choice allows one to drop the terms which are exponentially small \emph{i.e.} 
$\mathcal{O}(e^{-\epsilon u_0})$.  The leading sub-Planckian contributions to 
the Wightman function in this domain, after the removal of the regulator 
$\epsilon$, are given by
\begin{equation}\label{GsubTimelikeLong}
 G_{sub} = - \frac{1}{4\pi^2\Delta t^2}
\left[1 - \frac{2i~\delta_1~ \lstar }{\Delta t} 
+ \mathcal{O}(\lstar^2/\Delta t^2) \right] ~.
 \end{equation}
The super-Planckian contributions are exponentially small \emph{i.e.}
$G^{\epsilon}_{super} \sim e^{-\epsilon u_0}$. Therefore, the leading terms of 
the long-distance Wightman function with polymer corrections can be expressed as
\begin{equation}\label{GPolyTimelike}
G^{poly}(x,x') =  \frac{1}{4\pi^2\Delta x^2}
\left[1 - 2i \delta_{1} ~\sgn(\Delta t) \frac{\lstar}{|\Delta x|} 
+ \dots \right] ~,
\end{equation}
where $\sgn(\Delta t)$ is the \emph{signum} function and we have used $\Delta 
x^2=-\Delta t^2$ for the chosen frame of reference. The $2^{nd}$ term within 
the square bracket of the equation (\ref{GPolyTimelike}) signifies the leading 
correction that comes from the polymer quantization and contributes to the 
\emph{imaginary} part of the Wightman function. We should note here that in the 
limit $\lstar \to 0$, polymer corrected Wightman function (\ref{GPolyTimelike}) 
reduces to the Fock-space Wightman function.

\subsubsection{Short-distance Wightman function}

Similarly, one can also evaluate the short-distance Wightman function 
analytically in the domain $\Delta t \ll \lstar$. In this domain
the limit of integration $u_0 \sim (\Delta t/\lstar) \ll 1$. It is then
straightforward to compute the sub-Planckian contributions to the Wightman 
function as
\begin{equation}\label{GsubTimelikeShort}
 G_{sub}=\frac{1}{4\pi^2\lstar^2}\left[\frac{\tilde{g}_0^2}{2}
+ \frac{\delta_1 \tilde{g}_0^3}{3}
+ \mathcal{O}(\Delta t/\lstar) \right]  ~.
\end{equation}
The contributions from the super-Planckian modes to the 
Wightman function on the other hand can be expressed as
\begin{eqnarray}\label{GsuperTimelikeShort}
 G_{super}=\frac{1}{4\pi^2\lstar^2}\left[\frac{1}{16g_0}
-\frac{(1+i)}{16} \sqrt{\frac{\pi \Delta t}{\lstar}} 
+ \mathcal{O}(\Delta t/\lstar) \right] ~.
\end{eqnarray}
Therefore, the short-distance Wightman function with polymer corrections, 
including contributions from both sub-Planckian and super-Planckian modes,
can be written as
\begin{eqnarray}\label{GPolyTimelikeFinal}
 G^{poly}(x,x') && = \frac{1}{4\pi^2\lstar^2} \left[
\frac{1}{16g_0}\left(1 + 8 g_0^3 + \dots\right)  \right.  \nonumber \\
&& \left.  
- \frac{(1+i)}{16} \sqrt{\frac{\sgn(\Delta t) \pi |\Delta x|}{\lstar}}
+ \dots  \right] .
\end{eqnarray}
We note few important properties of the polymer corrected Wightman function
(\ref{GPolyTimelikeFinal}). Firstly, unlike in the case of Fock quantization, 
the Wightman function is bounded from above with the approximate maxima being at
$\sim 1/4\pi^2\lstar^2$. Secondly, the presence of inverse powers of $\lstar$ 
in the leading term implies that the modifications are \emph{non-perturbative} 
in nature. This asymptotic expression matches very closely to the numerical 
results as shown in the Fig.\ref{fig:Wightman}.

\subsection{Asymptotic forms of Wightman function for spacelike intervals}

Having computed the asymptotic Wightman function for timelike intervals we 
now focus on spacelike intervals. For a spacelike interval we can always 
choose a frame of reference where $\Delta t =0$. For simplicity, we choose such 
a frame and there the equation  (\ref{GPMDefinition}) reduces to the form
\begin{equation}
\label{GPMDefinitionSpacelike}
G_{\pm} =  \frac{i}{4\pi^2|\Delta\x|} \int dk ~k |c_k|^2
~e^{\mp i k |\Delta \x|} ~.
\end{equation}
The regulated expression for the corresponding Wightman function can then be 
expressed as
\begin{equation}\label{WightmanTimelikeSimplified}
G^{\epsilon} = \int du~h(u,|\Delta \x|) \left[e^{-iu} - e^{iu}\right]
e^{-\epsilon u}  ~,
\end{equation}
where the variable $u=k |\Delta \x|$ and the function $h(u,|\Delta \x|)$ is 
given by
\begin{equation}\label{hfunctionGeneralTimelike}
h(u,|\Delta \x|) =  \frac{i ~k~ |c_k|^2 }{4\pi^2 |\Delta \x|^2}  ~.
\end{equation}
The regulated expression for the sub-Planckian contributions to the Wightman 
function can be expressed as
\begin{equation}\label{GsubSpacelike}
 G^{\epsilon}_{sub} = \int_{0}^{u_0} du~h_{sub}(u,|\Delta \x|)~
\left[e^{-iu} - e^{iu}\right] e^{-\epsilon u} ~,
\end{equation}
where $u_0= g_0|\Delta \x|/\lstar $. The function $h_{sub}$ can be approximated 
as
\begin{equation}\label{hfunctionSpacelikeSubPlanckian}
h_{sub}(u,|\Delta \x|) = \frac{i}{8\pi^2 |\Delta \x|^2} 
\left[1 - \frac{2\delta_c \lstar u}{|\Delta \x|} \right]  ~.
\end{equation}
Similarly, the regulated super-Planckian contributions to the Wightman function 
can be expressed as
\begin{equation}\label{GsuperSpacelike}
 G^{\epsilon}_{super} = \int_{u_0}^{\infty} du~h_{super}(u,|\Delta \x|)
\left[e^{-iu} - e^{iu}\right] e^{-\epsilon u} ~,
\end{equation}
where the function  $h_{super}$ is given by
\begin{equation}\label{hfunctionSpacelikeSubPlanckian}
h_{super}(u,|\Delta \x|) = \frac{i~|\Delta \x|}{128\pi^2~\lstar^3} 
 ~u^{-3} ~.
\end{equation}
As earlier, the analytic evaluation of the equations (\ref{GsubSpacelike}) 
and (\ref{GsuperSpacelike}) are possible for asymptotically large or small 
intervals.

\subsubsection{Long-distance Wightman function}

In the long-distance ($|\Delta \x| \gg \lstar$) domain, the limit of 
integration $u_0 = g_0 (|\Delta \x|/\lstar) \gg 1$. Once again, we choose 
$(\epsilon u_0) \gg 1$ so that we can neglect terms which are exponentially 
small. The leading sub-Planckian contributions to the Wightman function in this 
domain are
\begin{equation}\label{GsubSpacelikeLong}
 G_{sub} =  \frac{1}{4\pi^2|\Delta \x|^2} \left[1 + 
\mathcal{O}\left(\lstar^2/|\Delta \x|^2\right) \right] ~.
\end{equation}
The contributions from the super-Planckian modes in the long-distance domain 
are exponentially small. Therefore, the leading terms of the long-distance 
Wightman function with polymer corrections can be expressed as
\begin{equation}\label{GPolySpacelikeLong}
G^{poly}(x,x') =  \frac{1}{4\pi^2\Delta x^2} \left[1 + 
\mathcal{O}\left(\lstar^2/\Delta x^2\right) \right] ~,
\end{equation}
where we have used $\Delta x^2 = \Delta \x^2$ for the chosen frame of reference.
Unlike for the case of timelike interval, the Wightman function \emph{does not} 
receive $\mathcal{O}\left(\lstar\right)$ corrections for spacelike interval 
from polymer quantization. In a sense this implies that the perturbative 
corrections terms from polymer quantization violate Lorentz invariance. 
Although, given the presence of a new dimension-full scale $\lstar$, the 
violation is expected.

\subsubsection{Short-distance Wightman function}

As earlier, we can compute the short-distance Wightman function analytically in 
the domain $|\Delta \x| \ll \lstar$. In this domain $u_0 = g_0 (|\Delta 
\x|/\lstar) \ll 1$. It is straightforward to evaluate the leading 
sub-Planckian contributions to the Wightman function which are given by
\begin{equation}\label{GsubTimelikeShort}
G_{sub}=\frac{1}{4\pi^2\lstar^2}\left[ \frac{g_0^2}{2}
- \frac{2\delta_c g_0^3}{3} 
+ \mathcal{O}(|\Delta \x|^2/\lstar^2) \right]  ~.
\end{equation}
Similarly, the leading contributions from the super-Planckian modes to 
the Wightman function can be expressed as
\begin{eqnarray}\label{GsuperTimelikeShort}
 G_{super} = \frac{1}{4\pi^2\lstar^2}\left[\frac{1}{16g_0}
-\frac{\pi}{64} \frac{|\Delta \x|}{\lstar}
+ \mathcal{O}(|\Delta \x|^2/\lstar^2) \right] ~.
\end{eqnarray}
The short-distance Wightman function with polymer corrections then can be 
written as
\begin{eqnarray}\label{GPolySpacelikeShort}
G^{poly}(x,x') = \frac{1}{4\pi^2\lstar^2}\left[
\frac{1}{16g_0}\left(1 + 8 g_0^3 + \dots\right)  \right. \nonumber \\
\left.  -\frac{\pi}{64} \frac{|\Delta x|}{\lstar} + \dots\right] ~.
\end{eqnarray}
Similar to the case of timelike intervals, the short-distance polymer corrected 
Wightman function for spacelike intervals remains bounded from above. The 
maximum magnitude of the Wightman function reaches approximately to the same 
value as for the case of the timelike intervals, when the separation interval is
taken to be zero.

\subsection{Emergence of effective spacetime discreteness from Wightman 
function}

We have already noted that once the integral regulator is removed the Wightman 
function (\ref{wightmanFock}) for a massless free scalar field in Fock 
quantization becomes inversely proportional to the invariant distance squared 
between the corresponding spacetime points. This property can be used 
as an effective probe to measure the spacetime distance between the two given 
points, as experienced by the scalar matter if we know the corresponding 
Wightman function. It can be made apparent by inverting the equation 
(\ref{wightmanFock}) to define an \emph{effective spacetime distance} between 
two given points $x$ and $x'$ as follows
\begin{equation}\label{EffectiveDistanceFromWightman}
\Delta x_{eff}^2 (x,x') \equiv \frac{1}{4\pi^2 ~|G(x,x')|}  ~.
\end{equation}
In the definition (\ref{EffectiveDistanceFromWightman}) we have used the 
magnitude of the Wightman function, as being a transition amplitude it is a 
complex-valued function in general. This limits the effective distance to probe 
only the magnitude of it. For Fock quantization $\Delta x_{eff}^2 (x,x') = 
|(-\Delta t^2+|\Delta \x|^2)|$ is the magnitude of usual invariant distance in 
the Minkowski spacetime. Therefore, when these two points $x,x'$ are taken 
infinitesimally close to each other then the effective distance between them 
vanishes as
\begin{equation}\label{EffectiveDistanceFromWightmanFockLimit}
\lim_{x\to x'} \Delta x_{eff}^2 (x,x')_{|Fock} = 0  ~.
\end{equation}
On the other hand, the Wightman function remains bounded from above in
polymer quantization for both timelike or spacelike intervals. So the 
analogous effective spacetime distance in the limiting case for polymer 
quantization can be written as
\begin{equation}\label{EffectiveDistanceFromWightmanPolymerLimit}
\lim_{x\to x'} \Delta x_{eff}^2 (x,x')_{|poly} \simeq \lstar^2 ~
 \left(\frac{64 \pi^2 g_0}{1 + 8g_0^3} \right) ~.
\end{equation}
Unlike in the case of Fock quantization, the equation 
(\ref{EffectiveDistanceFromWightmanPolymerLimit}) implies a non-vanishing value 
for the so called \emph{zero-point length}. The notion of zero-point length of 
the spacetime, such as seen in equation 
(\ref{EffectiveDistanceFromWightmanPolymerLimit}), has long been anticipated to 
arise from the possible quantum gravity effects quite generically 
\cite{Padmanabhan:1996ap,Smailagic:2003hm} as well as in specific approaches to 
quantum gravity such as in string theory 
\cite{Amati198941,Gross1988407,tamiaki}, non-commutative geometry 
\cite{Girelli:2004md,Douglas:2001ba}. However, we should emphasize that here we 
have applied polymer quantization only for the matter sector. In particular, the 
geometry has been treated classically. Therefore, the equation 
(\ref{EffectiveDistanceFromWightmanPolymerLimit}) appears to imply an 
\emph{emergence} of an effective discreteness in the spacetime. In this sense, 
the polymer quantization of matter field itself seems to capture certain aspects 
of quantum gravity effects.

\subsection{Numerical computation of Wightman function}

Analytical computation of the Wightman function in polymer quantization  does 
not appear to be possible for all possible spacetime intervals, unlike in Fock 
quantization. One could perform analytical computations only for the 
asymptotic regions. However, the asymptotic analysis does not present a reliable 
 picture for the intermediate regions. So in order to obtain a comprehensive 
picture of the Wightman function in polymer quantization we employ numerical 
techniques. Subsequently we evaluate the Wightman function for a wide range of 
spacetime intervals as permitted by the available computing resources.

\subsubsection{Matrix element $D_{\k}(\Delta t)$}

\begin{figure}[H]
 \includegraphics[scale=0.7]{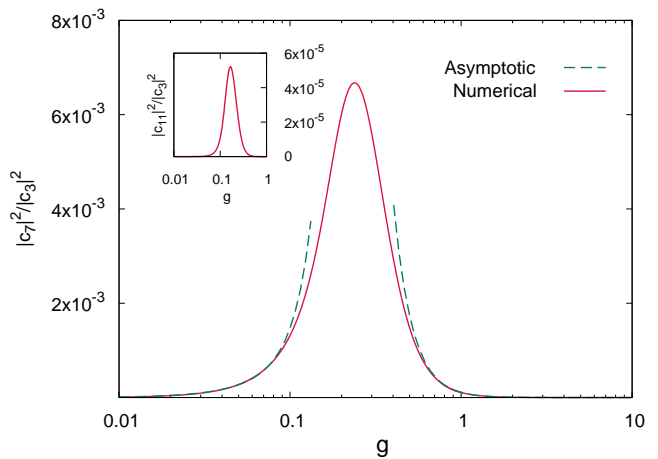}
 \caption{The solid line represents numerically evaluated ratio between 
$|c_7|^2$ and $|c_3|^2$. The dashed line represents the same ratio using
their respective asymptotic expressions. Both for the large $g$ and small 
$g$, compared to $g_0$, the asymptotes closely follow the numerical results. 
}
 \label{fig:c3c7comparison}
\end{figure}

It can be seen from the equation (\ref{GPMDefinition}) that the Wightman 
function is fully determined once the matrix element $D_{\k}(\Delta t)$ is 
known. In polymer quantization, the matrix element (\ref{DkFunctionGeneral}) can 
be expressed as
\begin{eqnarray}\label{Re-DkFunctionGeneral}
&&D_{\k}^{poly}(\Delta t) = \sum_{n}|c_{4n+3}|^2 e^{-i\Delta E_{4n+3}\Delta t},
\\ \nonumber
&=&|c_3|^2e^{-i\Delta E_3 \Delta t}\left[1+\frac{|c_7|^2}{|c_3|^2}
e^{-i(\Delta E_7-\Delta E_3) \Delta t}+ \dots\right] ~.
\end{eqnarray}
Unlike in Fock quantization where only the first term is non-vanishing in the 
summation, in polymer quantization there are infinitely many non-vanishing 
terms. From the FIG.\ref{fig:c3c7comparison}, we can see that  
$|c_7|^2/|c_3|^2\ll1$ even in the intermediate regions where $g \sim g_0$. 
In the asymptotic regions such behaviour can be seen analytically. It may 
also be shown that all other higher order coefficients are progressively 
smaller.

Therefore, in order to simplify the numerical computation we consider 
the contribution to the matrix element from $c_3$ term only. 

\subsubsection{Coefficient $c_k$ and energy gap $\Delta E_k$}

\begin{figure}[H]
\includegraphics[scale=0.75]{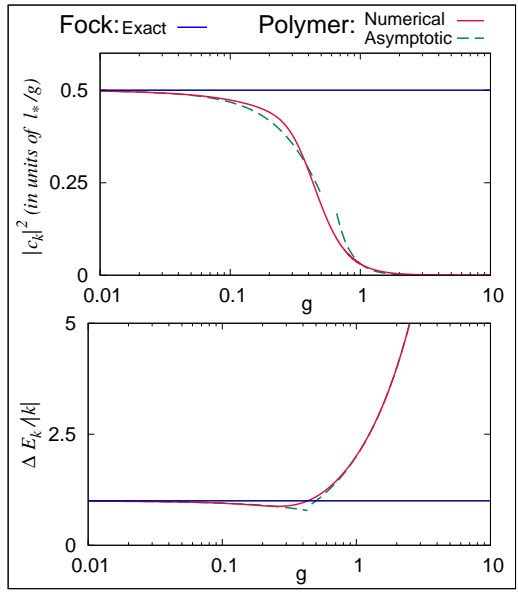}
\caption{In Fock quantization, $|c_k|^2 = \frac{1}{2} (\lstar/g)$ and it is 
represented as the solid blue line. The solid red line represents the numerical 
results and the dashed line represents the asymptotic expressions for $|c_k|^2$ 
in polymer quantization. 
In Fock quantization, $\Delta E_3/|\k|=1$, which is represented by the
straight solid line. The curved solid line represents the numerical results and 
the dashed line represents asymptotic expressions for polymer quantization. For 
both small and large values of $g$, compared to $g_0$, the asymptotic 
expressions closely follow the numerical results. 
\label{fig:ck} }
\end{figure}

In the Fock quantization the non-vanishing coefficient $c_1$ can be expressed
using dimensionless parameter $g$ as $|c_k|^2 \equiv |c_1|^2 = \frac{1}{2} 
(\lstar/g)$. For the  ease of comparison, we refer the coefficient $c_3$ as 
$c_k$ and the corresponding energy gap $\Delta E_3$ as $\Delta E_k$ also for 
polymer quantization.  In the FIG.\ref{fig:ck}, we have plotted the values 
$|c_k|^2$ as a function of the parameter $g$ using asymptotic expressions and 
numerically evaluated values.

In Fock quantization, the energy gap $\Delta E_k \equiv \Delta E_3 = |\k|$ 
hence the ratio $\Delta E_k/|\k|$ is unity for all values of $g$. However, in 
polymer quantization, that ratio $\Delta E_k/|\k|$ deeps below unity and has a 
minima at around $g \approx 0.26$. In the region where $g \gg g_0$, this ratio 
differs drastically (\ref{LargegDeltaE4n+3}) from the result of Fock 
quantization. We have plotted the behaviour of the energy gap $\Delta E_k$ as a 
function of $g$ in FIG.\ref{fig:ck}.

\subsubsection{Wightman function}

In order to perform the numerical evaluation, we scale the Wightman function as 
follows
\begin{equation}
G (\Delta t,\Delta \x) = \frac{1}{4\pi^2\lstar^2} ~\tilde{G}  ~,
\end{equation}
where $\tilde{G}$ is dimensionless. Using the equations 
(\ref{KGPropagatorDiffPM}) and (\ref{GPMDefinition}), the regulated 
expression of the \emph{scaled} and \emph{dimensionless} Wightman function 
can be expressed as
\begin{equation}\label{wightmantilde}
\tilde{G}^{\epsilon} = \int_{g_{min}}^{g_{max}}dg~ 
\tilde{h}(g,|\Delta \x|)~e^{-i u(g) - \epsilon g}.
\end{equation}
where $g_{min}$ and $g_{max}$ are approximate limits of integration which are 
used in numerical evaluation to represent $0$ and $\infty$ respectively. The 
function $u(g)$ can be expressed in terms of the dimensionless quantities as
\begin{equation}\label{ugFunction}
u(g) = g~\left(\frac{\Delta E_k}{|\k|}\right)
~ \left(\frac{\Delta t}{\lstar}\right) ~.
\end{equation}
Similarly, the function $\tilde{h}$ can be expressed in terms of dimensionless 
quantities as follows
\begin{equation}\label{ugFunction}
\tilde{h}(g,|\Delta \x|) = 2~ \left(\frac{|c_k|^2 g}{\lstar}\right) 
~ \left(\frac{\lstar}{|\Delta \x|}\right)
~ \sin\left(g \frac{|\Delta \x|}{\lstar} \right) ~.
\end{equation}
It may be noted that we have explicitly expressed spacetime intervals $\Delta 
\x$ and $\Delta t$  in the units of $\lstar$.

\begin{figure}[H]
 \includegraphics[scale=0.7]{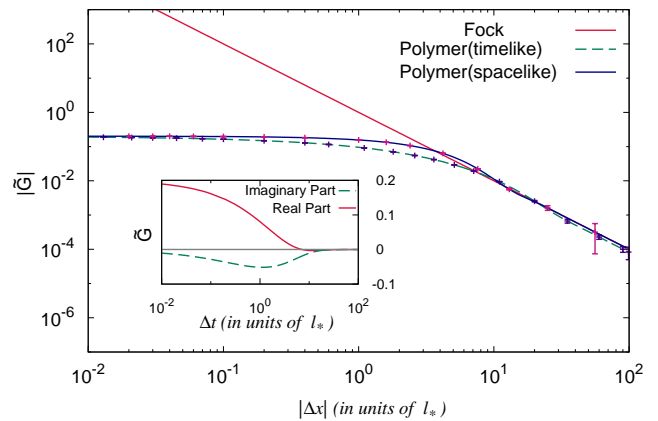}
 \caption{The Wightman function is scaled by a factor $1/4\pi^2\lstar^2$ and 
the solid red line represents the same in Fock quantization. The dashed green 
line represents absolute value of the scaled polymer Wightman function for 
timelike intervals and the solid blue line represents the same for spacelike 
intervals. For both timelike and spacelike intervals, the polymer Wightman 
function approaches the same value in the short distance. In large distance, 
Wightman function in polymer quantization mostly follows the result of the Fock 
quantization but in the short distance it differs significantly. The numerical 
error bars are \emph{amplified} by a factor $10^3$ to make them legible in the 
figure.
In the inset the solid red line represents the real part of the polymer 
Wightman function whereas the green dashed line represents the imaginary part.
}  
\label{fig:Wightman}
\end{figure}

From FIG.\ref{fig:Wightman}, we may see that numerically evaluated Wightman 
function in polymer quantization, differs significantly from Fock quantization 
in short-distance. However, for the large distance they closely follow each 
other. Furthermore, in the limit $\Delta x^2 \to 0$, numerically evaluated 
bounded value matches the asymptotic value $\tilde{G}_0 \approx 0.24$.

\subsubsection{Numerical errors and the domain}

The integrand in equation (\ref{wightmantilde}) involves exponential 
function. Therefore, by considering the limitation of double-precision
floating-point numbers, here we have used $g_{min}=10^{-3}$ ,
$g_{max}=10^3$ for numerical evaluation of the Wightman function. To ensure the 
convergence of the integration (\ref{wightmantilde}) within the desired 
precision, we choose an appropriate value of the regulator $\epsilon$
which depends on $g_{max}$ as well as on the chosen intervals. In particular, 
for a given set of spatial interval $\Delta \x$ and temporal interval $\Delta 
t$, the regulator should be chosen such that $(u(g)/g \pm |\Delta \x|/\lstar) > 
\epsilon$ and $(\epsilon ~g_{max}) >1$.

The numerical errors in computing Wightman function stems from the finite 
size of the divisions that we consider for integrating using trapezoidal rule.
Therefore, we can estimate this numerical errors by computing the same integral 
using two different division sizes and then considering their differences as 
follows
\begin{equation}\label{error}
 \delta G_{num} =|G|_{\Delta g_1}-|G|_{\Delta g_2} ~.
\end{equation}
Here we have considered $\Delta g_1 = 2\times10^{-4}$ and $\Delta g_2 = 
1\times10^{-4}$. It can be seen from FIG. \ref{fig:Wightman} that the numerical 
errors are much smaller than the evaluated values of the Wightman function for 
both the cases.

\section{Discussions}

In summary, we have studied the properties of the Wightman function \emph{i.e.} 
vacuum two-point function, corresponding to a massless free scalar field in
polymer quantization. In Fock quantization, the corresponding Wightman function 
 is inversely proportional to the invariant distance squared between the 
corresponding spacetime points. Therefore, the Wightman function diverges 
when these two points are taken to be infinitesimally close to each 
other. We have shown here that in contrast to the Fock quantization, Wightman 
function is bounded from above in polymer quantization. We have established 
this bounded nature of the Wightman function by asymptotic but analytic 
computation as well as using numerical methods. The bounded value of the 
polymer Wightman function is controlled by the polymer length scale $\lstar$ 
which is analogous to the Planck length.

We have argued that the Wightman function \emph{i.e.} vacuum two-point 
function, can be used as a probe to measure the effective spacetime distance 
between these two points, as experienced by the scalar field. In the case of 
polymer quantization, the bounded Wightman function leads to the notion of 
\emph{zero-point length} of spacetime. Such notion of the zero-point length has 
long been anticipated to arise from the possible quantum gravity effects  
quite generically \cite{Padmanabhan:1996ap,Smailagic:2003hm} as well as in 
specific approaches to quantum gravity such as in string theory 
\cite{Amati198941,Gross1988407,tamiaki}, non-commutative geometry 
\cite{Girelli:2004md,Douglas:2001ba}. However, in this article we have used 
polymer quantization only for matter field and the geometry has been treated 
classically. In this sense, the polymer quantization of matter field itself 
seems to capture certain aspects of quantum gravity effects. In other words, 
the property of the Wightman function in polymer quantization seems to 
imply an \emph{emergence} of an effective discreteness in the spacetime.

We now discuss few implications of this bounded nature of the Wightman 
function. In the study of \emph{Unruh effect}, the properties of the Wightman 
function plays an important role. In particular, the non-transient term of the 
instantaneous transition rate of the \emph{Unruh-DeWitt} detector contains a 
residue evaluated at the pole of the Wightman function. In contrast to the 
Fock quantization, there is no pole in polymer Wightman function as shown here.
Therefore, the non-transient term in the response function of the Unruh-Dewitt 
detector would disappear in polymer quantization \cite{Hossain:2016klt}. 
Nevertheless, there are some alternative views on the response of the 
Unruh-DeWitt detectors \cite{Kajuri:2015oza,Husain:2015tna}. The 
properties of the Wightman function as shown here also support the results of 
\cite{Hossain:2015xqa} where the violation of \emph{Kubo-Martin-Schwinger} 
condition is shown and criticism raised in \cite{Rovelli:2014gva} is addressed. 
The disappearance of the Unruh effect has also been seen using the method of 
Bogoliubov transformation \cite{Hossain:2014fma}. 
The bounded nature of the Wightman function also serves as an example where 
some aspects of the anticipation regarding quantum gravity to serve as a natural 
regulator \cite{Garay:1994en,Hossenfelder:2012jw,Hossenfelder:2005ed,
Hossenfelder:2004gj,Thiemann:1997rt,Kothawala:2013maa}, are realized.

We may also like to point out that the bounded nature of the Wightman function 
as shown here, is analogous to the behaviour of the effective Hubble parameter 
and the spectrum of inverse scale factor operator in loop quantum cosmology 
(LQC) \cite{Ashtekar:2006wn,Ashtekar:2006rx,Bojowald:2001xe,Bojowald:2001vw}. 
These results are also associated with some inverse powers of the 
distance scale, similar to the properties of the two-point function. In LQC, 
this crucial behaviour plays a key role in resolution of Big Bang singularity. 
However unlike in LQC, here we have applied polymer quantization only for scalar 
matter field rather than for the geometry.

\begin{acknowledgments}
We would like to thank Ritesh Singh for discussions. We thank Subhajit Barman
and Chiranjeeb Singha for their comments on the manuscripts. GS would like to
thank UGC for supporting this work through a doctoral fellowship. 
\end{acknowledgments}

%\bibliographystyle{apsrev}
%\bibliographystyle{aipauth4-1}

%\bibliography{bibtexfile}

\end{document}